\documentclass[a4paper,11pt]{article}
\usepackage{pos}
\usepackage{physics}
\usepackage{tikz-feynhand}

\newcommand{\Eq}[1]{Eq.~(\ref{#1})}
\newcommand{\alt}{\mathrel{\lesssim}}
\newcommand{\agt}{\mathrel{\gtrsim}}

\title{Axialvector diqaurk Mass and quark-diquark potential in $\Sigma_c$}

\author*[a]{Soya Nishioka}
\author[a]{Noriyoshi Ishii}

\affiliation[a]{RCNP, Osaka university,\\
1-11, Osaka Ibaraki Mihogaoka, Japan}

\emailAdd{nishioka@rcnp.osaka-u.ac.jp}
\emailAdd{ishiin@rcnp.osaka-u.ac.jp}

\abstract{The  axialvector  diquark is  studied  by  using 2+1  flavor
  Lattice QCD.
  Being a two-quark object, diquark has a non-neutral color charge.
  Hence  the two-point  correlators of  diquark fields  do not  have a
  particle pole due to the color confinement of QCD,
  and it is  not straightforward to study the diquark  mass by lattice
  QCD by using an exponential fit of a temporal two-point correlator.
  In order  to avoid this  difficulty, our  strategy is to  regard the
  diquark mass as a mass parameter of an effective quark-diquark model
  which is  constructed by using an  extended HAL QCD method  based on
  equal-time quark-diquark Nambu-Bethe-Salpeter (NBS) wave functions.
  We  attempt  to calculate  the  axial-vector  diquark mass  and  the
  quark-diquark potentials  between a charm quark  and an axial-vector
  diquark in the $\Sigma_c$ baryon.
  Lattice QCD  Monte Carlo calculation  is performed by using  the 2+1
  flavor QCD gauge configurations generated on $32^3\times 64$ lattice
  by PACS-CS Collaboration which corresponds to the pion mass of about
  700 MeV.
  As a result, a quark-diquark central potential of Cornell-type and a
  short-ranged spin-dependent potential are obtained.
  However,  from  a  quantitative  point  of  view,  the  gound  state
  convergence of the NBS wave functions  are not sufficient so that we
  obtain a  larger string tension  and a smaller  axial-vector diquark
  mass than we have expected phenomenologically.
  }

\FullConference{The 41st International Symposium on Lattice Field Theory (LATTICE2024)\\
 28 July - 3 August 2024\\
Liverpool, UK\\}

\begin{document}
\maketitle

\section{Introduction}

The singly charmed baryons, $\Lambda_{c}$ and $\Sigma_c^+$, consist of
one up quark,  one down quark and  one charm quark in  the quark model
picture.
They  have different  iso-spin  quantum numbers:  $\Lambda_{c}$ is  an
iso-scalar whereas $\Sigma_{c}$ is an iso-vector.
This difference  arises from  the distinct quantum  numbers of  the ud
clusters in $\Lambda_c$ and $\Sigma_c$.
The quantum  numbers of the  ud cluster in $\Lambda_c$  are $J^P=0^+$,
$I=0$  and  color  $\bf  \bar{3}$,   while  those  in  $\Sigma_c$  are
$J^P=1^+$, $I=1$ and color $\bf \bar{3}$.
The ud  cluster in  $\Lambda_c$ is  referred to  as the  scalar (good)
diquark, while  that in  $\Sigma_c$ is  called the  axial-vector (bad)
diquark.
The    mass   difference    between   $\Lambda_c$    and   $\Sigma_c$,
i.e. $m_{\Lambda_c}  < m_{\Sigma_c}$,  can be  attributed to  the mass
difference of these ud clusters.
In  fact, the  scalar  diquark  is expected  to  be  lighter than  the
axial-vector diquark,
as   suggested    by   the   color-magnetic   interaction    and   the
instanton-induced  interaction in  the  quark model,  which favor  the
scalar diquark.

Importance of diquarks extends beyond this charmed baryon example.
Diquarks  are  believed  to  play significant  roles  in  various  QCD
phenomena \cite{Jaffe:2005zz}.
However, studying diquark properties experimentally is challenging due
to QCD color confinement.
As  a result,  most known  diquark properties  remain qualitative  and
involve uncontrollable uncertainties,
since  they  are  inferred  not   from  direct  experiments  but  from
phenomenological arguments based on effective models.
One may wonder  whether lattice QCD, as a first-principle approach, can
provide an alternative means to study the diquarks.
However,  the  standard  lattice  QCD methods  cannot  directly  probe
diquarks due to color confinement.
Because two-point correlators lack a  particle pole in momentum space,
it  is impossible  to extract  a  diquark mass  using the  conventional
exponential fit to temporal two-point correlators.

Many efforts have been made to study diquark masses in lattice QCD,
with notable works classified into two main approaches:
(i)  Refs.   \cite{Hess:1998sd,  Babich:2007ah, Bi:2015ifa}  and  (ii)
Refs.     \cite{Alexandrou:2006cq,    Orginos:2005vr,    Green:2010vc,
  Francis:2021vrr}.
In  approach (i),  Landau  gauge  fixing is  employed  to compute  the
diquark two-point correlators,
after which the  standard hadron mass extraction method  is applied to
these gauge-fixed correlators.
While   their  results   align  with   naive  expectations   based  on
phenomenological  arguments, further  discussion  is needed  regarding
their consistency with color confinement.
In  approach (ii),  a static  quark  is introduced  to neutralize  the
diquark color charge,
and the energy of the total system is measured as the diquark mass.
However, the  interaction energy  between the  diquark and  the static
quark remains unsubtracted,
leading to an uncertainty of $O(\Lambda_{\rm QCD})$.

Recently, another method  is proposed in Ref.~\cite{Watanabe:2021nwe},
where, to avoid  the difficulty related to the  color confinement, the
scalar  diquark  mass   is  computed  as  a  mass   parameter  of  the
quark-diquark model  which is constructed  by an extension of  the HAL
QCD potential method \cite{ishii2007nuclear, aoki2010theoretical}.
%

The aim of our study is  to investigate the axial-vector diquark using
a similar method.
Specifically, by  applying an extended  HAL QCD potential  approach to
describe the $\Sigma_c$  baryon as a bound state of  a charm quark and
an axial-vector  diquark, we  aim to  determine both  the axial-vector
diquark mass and the quark-diquark potential.



\section{Formalism}

We begin  by recalling  that, since the  axial-vector diquark  has the
non-neutral  color charge  ($\bf \bar{3}$),  two-point correlators  of
diquark fields  do not possess  a particle-pole in the  momentum space
due to the color confinement of  QCD, which makes it impossible to use
a single-exponential fit in order to extract the diquark mass from the
temporal two-point correlators.
(See Sec. 60.6.3 in Ref.~\cite{ParticleDataGroup:2022pth}.)
We would  like to avoid  this problem in  order to obtain  the diquark
mass from the lattice QCD.
Our strategy is to  regard the diquark mass as a  mass parameter of an
effective  quark-diquark  model  which   is  constructed  by  HAL  QCD
potential    method   by    using    the   equal-time    quark-diquark
Nambu-Bethe-Salpeter (NBS) wave functions generated by lattice QCD.

To follow the  strategy, we consider the  equal-time quark-diquark NBS
wave  function for  $\Sigma_c^{++}$ in  the  center of  mass frame  in
Rarita-Schwinger form in the Coulomb gauge as
\begin{equation}
  \psi_{i\alpha}(\vec x - \vec y; J,M)
  \equiv
  \langle 0|
  D_{ai}(\vec x)
  q_{a\alpha}(\vec y)
  | \Sigma^{++}_{c}(J, M)\rangle,
\end{equation}
where     $|0\rangle$      denotes     the     QCD      vacuum     and
$|\Sigma^{++}_{c}(J,M)\rangle$ denotes  the ground state for  the even
parity $\Sigma_c^{++}$ baryon in the rest frame with $J$ and $M$ being
the total angular momentum and the magnetic quantum number.
In this  paper, we  restrict ourselves  to the  case with  $J=1/2$ and
$3/2$.
$q_{a\alpha}(y)$ denotes  the Dirac spinor  field for the  charm quark
with the color index $a=1,2,3$  and the Dirac index $\alpha=1,2$ which
is restricted  to the upper  components in the  Dirac non-relativistic
representation.
$D_{ai}(x)$ denotes the composite axial-vector diquark field with $i =
1,2,3$ being  the Lorentz  index restricted  to the  spatial subspace,
which is defined as
\begin{equation}
  D_{ai}(x)
  \equiv
  \epsilon_{abc}\,
  u_{b}^{T}(x) C\gamma_i u_{c}(x),
\end{equation}
where  $u(x)$  denotes  the  u  quark field  and  $C  \equiv  \gamma_4
\gamma_2$ denotes the charge conjugation matrix.

The equal-time  NBS wave  functions are  related to  the quark-diquark
four-point correlator in the positive $t$ region as
\begin{eqnarray}
  C_{i\alpha}(\vec x, \vec y, t; \mathcal{J})
  &\equiv&
  \langle 0 |
  T\left[
  D_{ai}(\vec x, t) q_{a\alpha}(\vec y, t)
  \cdot
  \mathcal{J}(t=0)
  \right]
  |0\rangle
  \\\nonumber
  &=&
  \sum_{n}
  \langle 0 | D_{ai}(\vec x) q_{a\alpha}(\vec y)| n\rangle
  \cdot
  e^{-E_n t}
  \langle n| \mathcal{J}|0\rangle,
\end{eqnarray}
where $|n\rangle$ denotes the $n$-th eigenstate of the Hamiltonian and
$E_n$ denotes the eigenenergy.
$\mathcal{J}$ denotes the source operator.
In  this paper,  we restrict  ourselves to  the wall  source operators
defined by
\begin{equation}
  \mathcal{J}_{J,M}
  \equiv
  \sum_{\vec x,\vec y,\vec z}
  \bar{q}_{a\alpha}(\vec x)
  \cdot
  \epsilon_{abc}
  \bar{u}_b(\vec y) C\gamma_i \bar{u}^T_c(\vec z)
  \cdot
  (1, i; 1/2, \alpha|J,M),
\end{equation}
where  the  last factor  in  the  r.h.s.  denotes  the  Clebsch-Gordan
coefficients  $(j_1=1,  m_1  =  i;  j_2  =  1/2,  m_2  =  \alpha|J,M)$
associated with  the decomposition  ${\bf 1}\otimes  {\bf 1/2}  = {\bf
  1/2}\oplus{\bf 3/2}$.
Hence, in the large $t$ limit, the ground-state NBS wave functions for
$\Sigma_c$ baryon  in the spin  $J$ channels are obtained as
\begin{equation}
  \psi_{i\alpha}(\vec r \equiv \vec x - \vec y; J,M)
  \propto
  C_{i\alpha}(\vec  x,  \vec y,  t; \mathcal{J}_{J,M}).
\end{equation}

We  demand  that  the  quark-diquark NBS  wave  function  satisfy  the
Schr\"odinger equation as
\begin{equation}
  \label{eq:schroedinger-eq}
  \left(
  \hat H_0 + \hat V
  \right)
  \psi(\vec r; J,M)
  =
  \left( M_J - m_q - m_D \right)
  \psi(\vec r; J,M),
\end{equation}
where $\hat H_0 \equiv - \nabla^2/(2\mu)$ denotes the kinetic operator
with $\mu \equiv \frac1{1/m_q +  1/m_D}$ being the reduced mass.
$m_q$  and $m_D$  denote the  charm  quark mass  and the  axial-vector
diquark mass, respectively.
For the time  being, we proceed our argument treating  $m_q$ and $m_D$
as unknown parameters, which will be determined later.
Note  that  ${\mathcal E}_J  \equiv  M_J  -  m_q  - m_D$  denotes  the
``binding energy''  of this system  with $M_J$  being the mass  of the
ground state $\Sigma_{c}^{++}$ baryon of total angular momentum $J$.
$\hat V$ denotes the quark-diquark potential, which is truncated as
\begin{equation}
  \hat V
  \simeq
  V_0(\vec r)
  +
  V_{\rm s}(\vec r)\,\mathbf{s}_q \cdot \mathbf{s}_D
  + \cdots,
\end{equation}
where $V_0(\vec r)$ and $V_{\rm s}(\vec r)$ denote the central and the
spin-dependent  potential,   respectively,  with   $\mathbf{s}_q$  and
$\mathbf{s}_{D}$ being the spin operators  for the charm quark and the
axial-vector diquark, respectively.
By  using  $\mathbf{s}_D \cdot  \mathbf{s}_q  =  -1$ for  $J=1/2$  and
$\mathbf{s}_D \cdot \mathbf{s}_q = +1/2$ for $J=3/2$,
the Schr\"odinger equation \Eq{eq:schroedinger-eq} splits into $J=1/2$
and $3/2$ channels as
\begin{eqnarray}
  \left(
  \hat H_0 + V_0(\vec{r}) -V_s(\vec{r})
  \right)
  \psi_{1/2}(\vec{r})
  &=&
  (M_{1/2}-m_q-m_D)
  \psi_{1/2}(\vec{r})
  \\\nonumber
  \left(
  \hat H_0 + V_0(\vec{r}) + (1/2) V_s(\vec{r})
  \right)
  \psi_{3/2}(\vec{r})
  &=&
  (M_{3/2}-m_q-m_D)
  \psi_{3/2}(\vec{r}),
\end{eqnarray}
where  we introduce  a  short-hand notation  $\psi_{J}(\vec r)  \equiv
\psi_{i\alpha}(\vec r; J,M)$ for notational simplicity.
These equations are solved for $V_0(\vec r)$ and $V_{\rm s}(\vec r)$ as
\begin{eqnarray}
  \label{eq:potentials}
  V_0(\vec{r})
  &=&
  \frac{1}{3}
  \left(2 M_{3/2} + M_{1/2}\right)
  - m_q
  - m_D
  + \frac1{2\mu}
  \left(
  \frac{2}{3}
  \frac{
    \nabla^2\psi_{3/2}({\vec{r}})
  }{
    \psi_{3/2}({\vec{r}})
  }
  +\frac{1}{3}
  \frac{
    \nabla^2\psi_{1/2}({\vec{r}})
  }{
    \psi_{1/2}({\vec{r}})
  }
  \right)
  \\\nonumber
  V_{\rm s}({\vec{r} })
  &=&
  \frac{2}{3}
  \left(M_{3/2}-M_{1/2}\right)
  -\frac{1}{3\mu}
  \left(
  \frac{
    \nabla^2\psi_{3/2}({\vec{r} })
  }{
    \psi_{3/2}({\vec{r} })
  }
  -\frac{
    \nabla^2\psi_{1/2}({\vec{r} })
  }{
    \psi_{1/2}({\vec{r} })
  }
  \right).
\end{eqnarray}
These relations express $V_0(\vec r)$ and $V_{\rm s}(\vec r)$ by using
$\psi_{J}(\vec r)$, $M_J$, $m_q$ and $m_D$ as inputs,
where  we note  that  $m_q$  and $m_D$  are  treated  as unknown  mass
parameters which  should not be  determined by the standard  method of
the exponential fit.

Kawanai  and  Sasaki encountered  the  similar  problem in  $c\bar  c$
sector,
where they proposed  a prescription to determine the  charm quark mass
in a self-consistent manner with HAL QCD method \cite{Kawanai:2011xb}.
In  their approach,  they  require that  the spin-dependent  potential
should vanish at long distance.
Following  their  prescription, we  impose  the  condition as
\begin{equation}
  V_{\rm s}(\vec{r}) \to 0
  \hspace{2em}
  \mbox{as}
  \hspace{2em}
  r \to \infty,
\end{equation}
which leads us to
\begin{equation}
  \mu = -\lim_{r \to \infty} F_{\rm KS}(\vec r)
  \label{eq:reduced-mass}
\end{equation}
where we  introduce Kawanai-Sasaki  function $F_{\rm KS}(\vec  r)$ for
later convenience as
\begin{equation}
  F_{\rm KS}(\vec r)
  \equiv
  \frac1{2(M_{3/2}-M_{1/2})}
  \left(
  \frac{
    \nabla^2 \psi_{3/2}({\vec{r}})
  }{
    \psi_{3/2}({\vec{r}})
  }
  -\frac{
    \nabla^2\psi_{1/2}({\vec{r}})
  }{
    \psi_{1/2}({\vec{r}})
  }
  \right).
  \label{eq:FKS}
\end{equation}
The reduced mass $\mu$ in \Eq{eq:reduced-mass} together with the charm
quark mass $m_q$ obtained by  using the Kawanai-Sasaki prescription in
the $c\bar{c}$ sector, the axial-vector diquark mass is obtained as
\begin{equation}
  m_D = \frac{1}{1/\mu - 1/m_c}.
  \label{eq:diquark-mass}
\end{equation}


\section{Numerical results}

\subsection{Lattice QCD setup}
We employ the  2+1 flavor QCD gauge configurations  on $32^3\times 64$
lattice generated by PACS-CS Collaboration \cite{PACS-CS:2008bkb}.
These gauge configurations are generated  by employing the RG improved
Iwasaki  gauge action  at $\beta  = 1.90$  and the  non-perturbatively
$O(a)$-improved  Wilson quark  (clover) action  at $\kappa_{\rm  ud} =
0.13700$ and $\kappa_{\rm s} = 0.13640$ with $C_{\rm SW} = 1.715$.
This parameter set corresponds to the lattice spacing $a = 0.0907(13)$
fm ($a^{-1} = 2176(31)$ MeV), the spatial extent $L = 32a \simeq 2.90$
fm.
The  charm  quark  is  introduced by  the  quenched  approximation  by
employing the relativistic heavy quark  (RHQ) action, for which we use
the parameters given in Ref.~\cite{PACS-CS:2011ngu}.
Two-point and four-point correlators are calculated by using the quark
propagators which are obtained with  the wall source after the Coulomb
gauge fixing is applied to the gauge configurations.
Statistics  is  improved  by  using 64  source  points  by  temporally
shifting the gauge configurations.
The time-reversal and the charge  conjugation symmetries are also used
to double the statistics.
Statistical  errors  are  estimated with  the  Jackknife  prescription
employing the bin size of 20 configurations.
Some  reference hadron  masses obtained  by  this setup  are given  as
follows:   $m_{N}  \simeq   1583$   MeV,   $m_{\pi}\simeq  702$   MeV,
$m_{\eta_c}\simeq   3025$   MeV,   $m_{J/\psi}   \simeq   3144$   MeV,
$m_{\Lambda_c^{1/2+}}\simeq  2690$  MeV,  $m_{\Sigma_c^{1/2+}}  \simeq
2777$ MeV and $m_{\Sigma_c^{3/2+}} \simeq 2859$ MeV.

\subsection{Results}
\begin{figure}[h]
    \centering
    \includegraphics[width=0.9\linewidth]{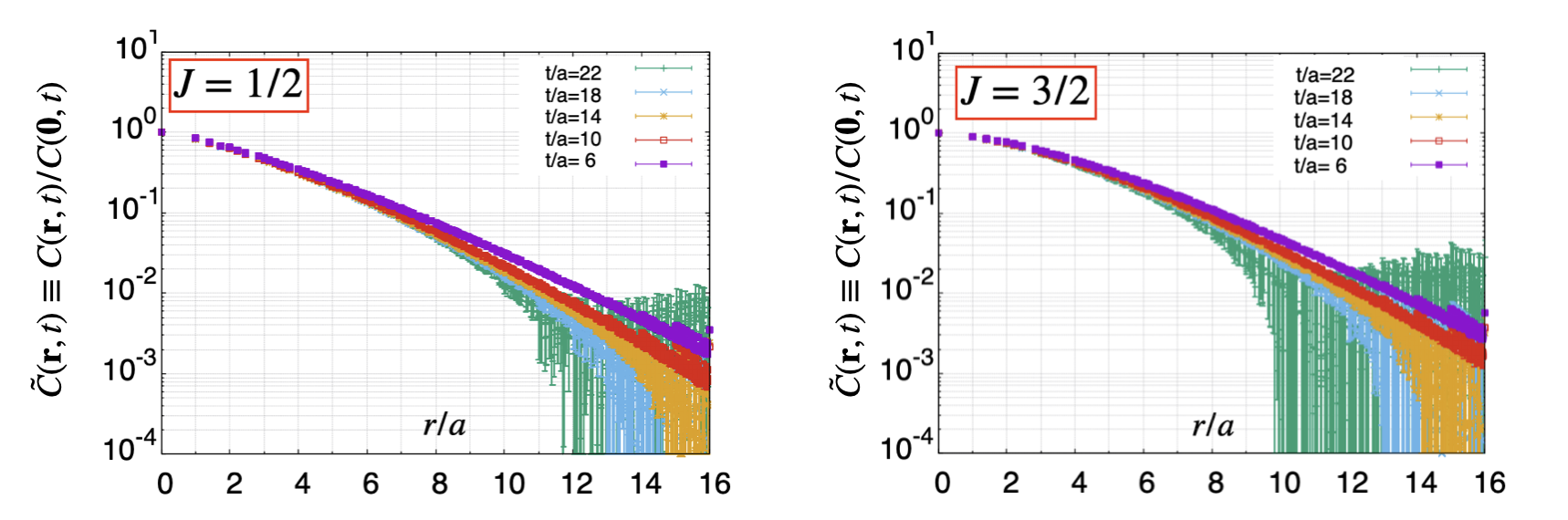}
    \caption{Normalized  four-point correlators  $\tilde C(\vec  r, t)
      \equiv C(\vec r, t)/C(\vec 0, t)$ for $J=1/2$ (left) and $J=3/2$
      (right).}
    \label{fig:4-point}
\end{figure}

The results of four-point correlators are shown in Figure \ref{fig:4-point},
where the normalized four-point correlator $\tilde C(\vec r, t) \equiv
C(\vec r, t)/C(\vec  0, t)$ is plotted  against $r$ for $t/a  = 6, 10,
14, 18, 22$.
We see that  the convergence is quicker in the  short distance region,
while it is slower in the long distance region.
If we  restrict ourselves to  the spatial  region $r/a \alt  7$, rough
convergence is achieved at $t/a = 18$.
(For $r/a  \agt 7$,  the convergence becomes  gradually worse  and the
error bar becomes larger.)
%
In  what follows,  we accept  the 4-point  correlators at  $t/a=18$ as
roughly converged NBS wave functions, and proceed our calculations.

\begin{figure}[t]
    \centering
    \includegraphics[width=0.4\linewidth]{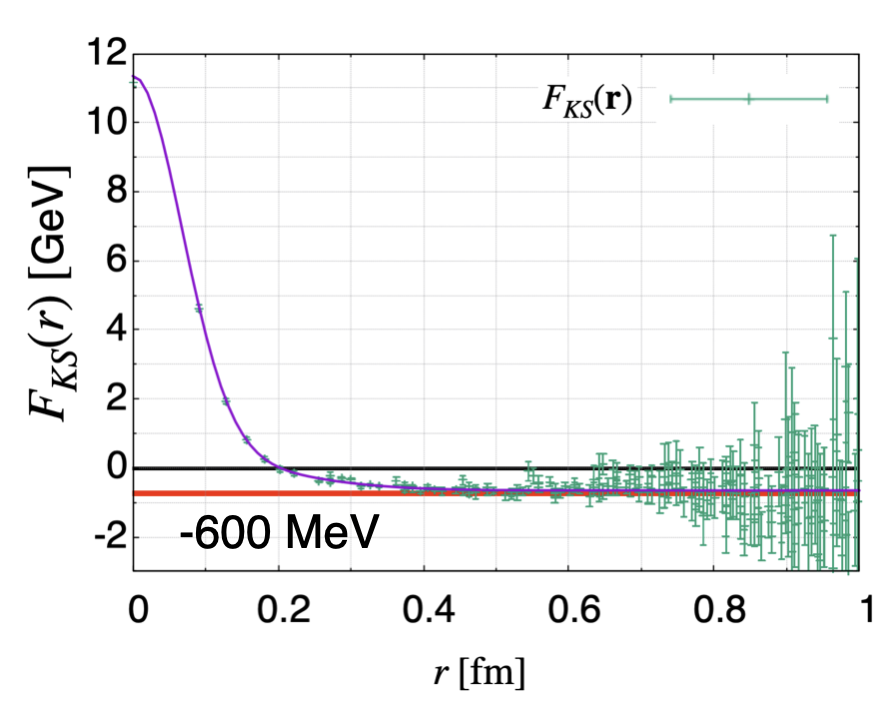}
    \caption{Kawanai Sasaki function $F_{\rm KS}(r)$ for $t/a = 18$.}
    \label{fig:FKS}
\end{figure}

Figure  \ref{fig:FKS}  shows  the   plot  of  Kawanai-Sasaki  function
$F_{KS}(r)$ against $r$ for $t/a = 18$.
The purple  curve denotes the  result of the 2-Gaussian  fit employing
the functional form:
$f(r)
\equiv
A \exp(-B r^2) + C\exp(-D r^2) + E,
$
where $A,B,C,D$ and $E$ are used as fit parameters.
Since the long distance limit  of $F_{\rm KS}(r)$ is obtained from
the constant part of the fit function,
we obtain the reduced mass:
$\mu = -\lim_{r\to\infty} F_{\rm KS}(r) = -E \simeq 600\,\mbox{MeV}$.
By applying  the Kawanai-Sasaki prescription  to the $c\bar{c}$  sector, we
obtain the charm quark mass $m_q \simeq 1950$ MeV.
By using \Eq{eq:diquark-mass}, we obtain the axial-vector diquark mass
$m_D \simeq 867$  MeV, which is significantly smaller  than the scalar
diquark    mass    $m_{SD}    \simeq    1273$    MeV    obtained    in
Ref.~\cite{Watanabe:2021nwe},
where  the same  gauge  configurations  with the  same  pion mass  are
employed.
The main reasons for this underestimate  seem to be the following two:
(i)  ground-state convergence  of the  four-point correlators  are not
sufficient at  long distance, which  results in an uncertainty  in the
evaluation of the constant part of $F_{\rm KS}(r)$.
(ii)  Ambiguity in  the charm  quark  mass.
In  Ref.~\cite{Watanabe:2021nwe},  significantly smaller  charm  quark
mass $m_q  \simeq 1686$  MeV is  used than our  charm quark  mass $m_q
\simeq 1950$ MeV.
One reason  for this discrepancy is  that Ref.~\cite{Watanabe:2021nwe}
uses the odd  parity $c\bar{c}$ spectrum to determine  the charm quark
mass instead of Kawanai-Sasaki prescription.
Another reason  is that  fit of this  $c\bar{c}$ central  potential is
technically involved.
In order  to perform a quantitative  fit in the whole  spatial region,
the simple Cornell-type functional form is not enough,
and    at    least    $\log(r)$    term    has    to    be    included
\cite{Watanabe:2021nwe}.
In  addition,  at short  distance,  the  violation of  the  rotational
symmetry is severe.

\begin{figure}[t]
    \centering
    \includegraphics[width=0.9\linewidth]{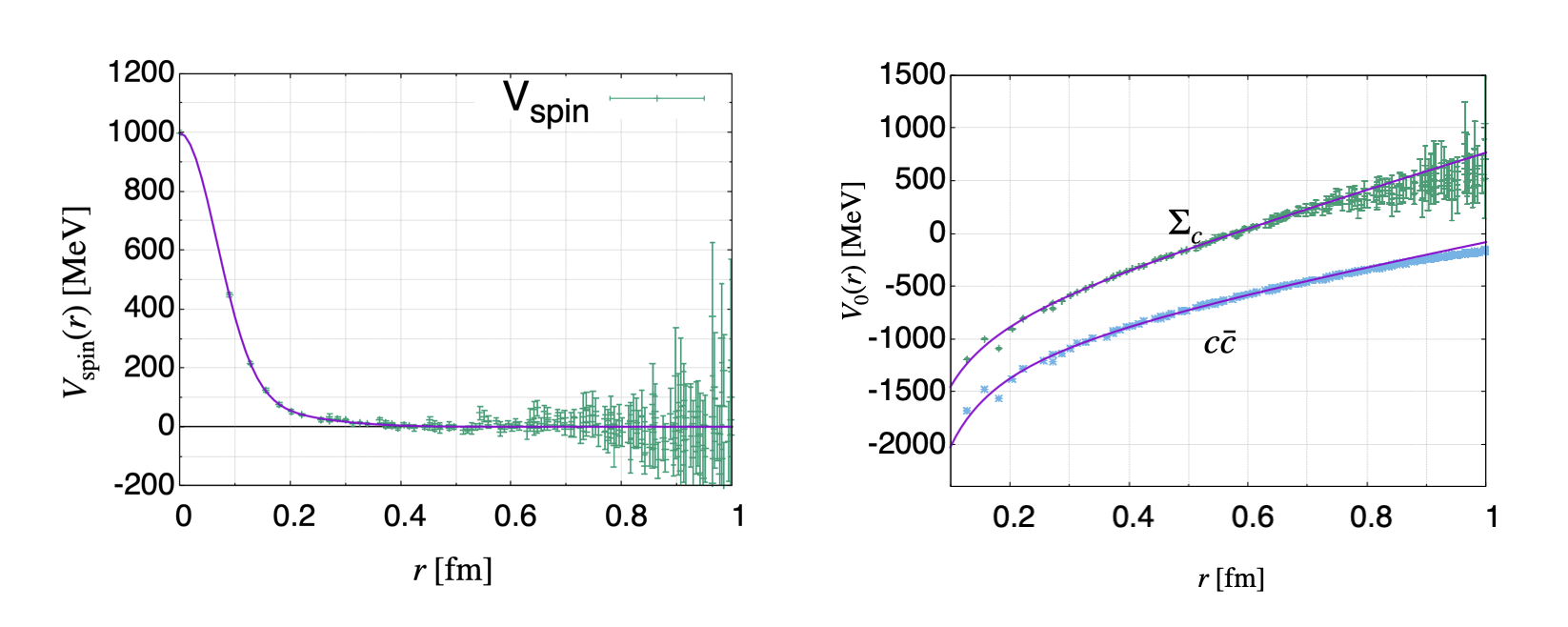}
    \caption{The spin dependent quark-diquark potential $V_{\rm s}(r)$
      (left)  and  the   central  quark-diquark  potential  $V_{0}(r)$
      together with the central $c\bar{c}$ potential (right).}
    \label{fig:potential}
\end{figure}

Figure \ref{fig:potential} shows the quark-diquark potentials obtained
from \Eq{eq:potentials}  with $m_q  \simeq 1950$  MeV and  $m_D \simeq
867$ MeV.
We  see that  the  spin-dependent potential  $V_{\rm  s}(r)$ is  short
ranged and  that the  quark-diquark central  potential $V_0(r)$  is of
Cornell-type:
$
  V_{\rm Cornell}(r)
  \equiv
  -A/r + \sigma r + \mbox{const}.
$
For   comparison,   $c\bar{c}$   potential    is   added   in   Figure
\ref{fig:potential}. 
These two central potentials are fitted with $V_{\rm Cornell}(r)$,
which leads to $A \simeq 86$ MeV fm and $\sqrt{\sigma} \simeq 565$ MeV
for quark-diquark sector, and $A \simeq 103$ MeV fm and $\sqrt{\sigma}
\simeq 459$ MeV for $c\bar{c}$ sector.
We see that, at long  distance, the quark-diquark central potential is
a bit steeper  than the $c\bar{c}$ central potential, which  may be an
artifact caused by the underestimate of the reduced mass $\mu$ through
the overall factor in \Eq{eq:potentials}.

For  quantitative calculation  of the  axial-vector diquark  mass, the
most important thing is to improve the ground-state convergence of the
quark-diquark four-point correlator.
For this  purpose, we plan  to use  the time-dependent HAL  QCD method
\cite{Ishii:2012ssm} and the variational  method to improve the source
operators in the near future.

\section{Conclusion}
We have  studied the axial-vector  diquark mass and  the quark-diquark
potentials between an axial-vector diquark  and a charm quark by using
2+1 flavor lattice QCD.

Since the axial-vector diquark has non-neutral color charge ($\bf \bar
3$), the  two-point correlator  does not have  a particle-pole  in the
momentum space due to the color confinement of QCD,
so  that the  standard  method  to calculate  the  hadron  mass by  an
exponential fit of the temporal  two-point correlator is not usable to
obtain the diquark mass.
In  this paper,  to avoid  this difficulty,  we have  resorted to  the
Kawanai-Sasaki extension of the HAL QCD potential method,
where  the  diquark mass  and  quark-diquark  potentials are  obtained
simultaneously
%
by demanding  that the  spin-dependent quark-diquark  potential should
vanish in the long distance limit.

Numerical calculations have been performed by using the 2+1 flavor QCD
gauge configurations  generated on $32^3\times 64$  lattice by PACS-CS
Collaboration employing the RG-improved Iwasaki gauge action at $\beta
=  1.90$  and  the  non-perturbatively  $O(a)$-improved  Wilson  quark
(clover)  action at  $\kappa_{\rm  ud}=1.3700$ and  $\kappa_{\rm s}  =
1.3640$ with $C_{\rm SW} = 1.715$.
The charm quark has been  incorporated with the quenched approximation
by using the relativistic heavy quark (RHQ) action.
The setup  lead to the lattice  spacing $a^{-1} = 2176(31)$  MeV ($a =
0.0907(13)$ fm) and the pion mass $m_{\pi} \simeq 702$ MeV.

By using the charm quark mass $m_q \simeq 1950$ MeV which was obtained
by applying the Kawanai-Sasaki extension of  the HAL QCD method to the
$c\bar{c}$  sector, we  have  obtained the  axial-vector diquark  mass
$m_{D} \simeq 867$ MeV, which is significantly smaller than the scalar
diquark mass  $m_{SD} \simeq  1273$ MeV obtained  by a  similar method
with     exactly     the      same     gauge     configurations     in
Ref.~\cite{Watanabe:2021nwe}.
The underestimate  of the axial-vector  diquark mass has seemed  to be
mainly  due  to  the  insufficient  ground-state  convergence  of  the
quark-diquark  four-point correlators  in  the  long spatial  distance
region.

We  have  also  obtained   the  quark-diquark  potentials  between  an
axial-vector diquark and a charm quark.
We have obtained the spin-dependent  potential $V_{\rm s}(r)$ which is
of  short-ranged and  the  central potential  $V_{0}(r)$  which is  of
Cornell-type.
However, we have seen that the long distance behavior of $V_{0}(r)$ is
steeper  than that  of  the central  $c\bar{c}$  potential, which  has
seemed  to be  due to  the underestimate  of the  axial-vector diquark
mass.

In the future, we will try  to improve the ground-state convergence of
the quark-diquark  four-point correlators by using  the time-dependent
HAL  QCD method.
We  will  also try  to  improve  the  source  operators by  using  the
variational method.
These improvements are important in making a precise comparison of the
axial-vector diquark and the scalar diquark.
We are interested in the quark mass dependence of our results, because
the calculations in this paper have  been carried out by employing the
gauge configurations which correspond to rather heavy pion mass.

\acknowledgments{Lattice QCD calculations have  been done by using the
  supercomputer  SQUID  at  the  Cyber Media  Center  (CMC)  of  Osaka
  University under the support of  Research Center for Nuclear Physics
  (RCNP) of Osaka University.
  We thank PACS-CS Collaboration and  JLDG/ILDG for the 2+1 flavor QCD
  gauge configurations.
  We also thank the  lattice QCD library bridge++ \cite{Ueda:2014rya},
  a modified version of which is used for our calculation.
  This work was  supported by JST SPRING, Grant  Number JPMJSP2138 and
  JSPS KAKENHI Grant Number JP21K03535.  }

\newpage

\bibliography{main.bib}
\bibliographystyle{unsrt}

\end{document}